\documentclass[prx,showpacs,twocolumn, preprintnumbers,
amsmath,amssymb,superscriptaddress,nofootinbib,longbibliography,nofootinbibjustification=justified,singlelinecheck=false]{revtex4-2}

\usepackage{amsfonts}
\usepackage{graphicx}
\usepackage{hyperref} 
\usepackage{cleveref}
\hypersetup{
    hidelinks,
    colorlinks=true,
    breaklinks=true,
    citecolor=SteelBlue,
    filecolor=LimeGreen,
    linkcolor=MediumBlue,
    urlcolor=MediumPurple,
    pdfauthor={Xuemei Gu, Mario Krenn}
}
\usepackage[svgnames]{xcolor}
\usepackage{xspace}
\usepackage{multirow}
\usepackage{array}
\usepackage{mathtools}
\usepackage{multirow}

\renewcommand{\arraystretch}{1.}  
\begin{document}
\title{Forecasting high-impact research topics\\via machine learning on evolving knowledge graphs}

\author{Xuemei Gu}
\email{xuemei.gu@mpl.mpg.de}
\author{Mario Krenn}
\email{mario.krenn@mpl.mpg.de}
\affiliation{ Max Planck Institute for the Science of Light, Staudtstrasse 2, 91058 Erlangen, Germany}

\begin{abstract}
The exponential growth in scientific publications poses a severe challenge for human researchers. It forces attention to more narrow sub-fields, which makes it challenging to discover new impactful research ideas and collaborations outside one's own field. While there are ways to predict a scientific paper's future citation counts, they need the research to be finished and the paper written, usually assessing impact long after the idea was conceived. Here we show how to predict the impact of onsets of ideas that have never been published by researchers. For that, we developed a large evolving knowledge graph built from more than 21 million scientific papers. It combines a semantic network created from the content of the papers and an impact network created from the historic citations of papers. Using machine learning, we can predict the dynamic of the evolving network into the future with high accuracy (AUC values beyond 0.9 for most experiments), and thereby the impact of new research directions. We envision that the ability to predict the impact of new ideas will be a crucial component of future artificial muses that can inspire new impactful and interesting scientific ideas.
\end{abstract}

\maketitle

\section*{Introduction}

As we see an explosion in the number of scientific articles
\cite{fortunato2018science,wang2021science,bornmann2021growth,krenn2023forecasting}, it becomes increasingly challenging for researchers to find new impactful research directions beyond their own expertise. Consequently, researchers might have to focus on narrow subdisciplines. A tool that can read and intelligently act upon scientific literature could be an enormous aid to individual scientists in choosing their next new and high-impact research project, which -- on a global scale -- could significantly accelerate science itself.

These days, a natural first choice for an AI-assistant would be powerful large-language-models (LLMs) such as GPT-4 \cite{achiam2023gpt}, Gemini \cite{team2023gemini}, LLaMA-2 \cite{touvron2023llama} or custom-made models \cite{wang2023learning}. However, these models often struggle in scientific reasoning, and it remains unclear how they can suggest new scientific ideas or evaluate their impact in a reliable way in the near term.

An alternative and complementary approach is to build scientific semantic knowledge graphs. Here, the nodes represent scientific concepts and the edges are formed when two concepts are researched together in a scientific paper \cite{wang2021science}. While this approach extracts only small amounts of information from each paper, surprisingly non-trivial conclusions can be drawn if the underlying dataset of papers is large. An early example of this is a work in biochemistry \cite{rzhetsky2015choosing}. The authors use their semantic network, where nodes represent biomolecules, to find new potentially more efficient exploration strategies for the bio-chemistry community on a global scale. In these semantic networks, an edge between two concepts indicates that researchers have jointly investigated these research concepts. The edges are drawn from papers, thus they are created at a specific time when the paper was published. In this way, one creates an evolving semantic network that captures what researchers have investigated in the past. With such an evolving network, one can ask how the network might evolve in the future. In the scientific context, this question can be reformulated into what scientists will research in the future. For example, if two nodes do not share an edge, one can ask whether they will share an edge in the next three years -- or, alternatively, whether scientists will investigate these two concepts jointly within three years. This question, denoted as a link-prediction problem in network theory \cite{martinez2016survey}, has been successfully demonstrated with high prediction quality for semantic networks in the field of quantum physics \cite{krenn2020predicting} and artificial intelligence \cite{krenn2023forecasting}. These works focus on the question \textit{what scientists will work on}, completely leaving out which of these topics will be impactful. 

Impact in the scientific community is often approximated (for lack of better metrics \cite{nature2017challenge,nature2024facets}) by citations \cite{fortunato2018science, wang2021science, barabasi2018formula, frank2019evolution}, including exciting results that find interpretable mathematical models to describe citation evolution \cite{wang2013quantifying, ke2015defining, sinatra2016quantifying, wu2019large}. Beside concrete mathematical modelling, impact of scientific papers has also been predicted using advanced statistical and machine-learning methods that use meta-data such as including authors and affiliations \cite{weis2021learning}, the content and the references of the paper \cite{bai2017overview,xia2023review}. Techniques employed for the predictions of individual paper impact using a combination of characteristics include support-vector machines \cite{fu2010using}, regression \cite{yu2014citation, stegehuis2015predicting, weihs2017learning}, dense \cite{ruan2020predicting} or graph neural networks \cite{he2023h2cgl}.

The prediction of a paper's impact however is only possible after the research is completed, and long after its underlying idea is created. A true scientific assistant or muse however should contribute at the earliest stage of the scientific cycle, when the idea for the next impactful research project is born. One solution is the prediction at the concept level. Specifically, we can ask the question \textit{Which scientific concepts, that have never been investigated jointly, will lead to the most impactful research?}.

\begin{figure*}[!t]
    \centering
    \includegraphics[width=0.85\linewidth]{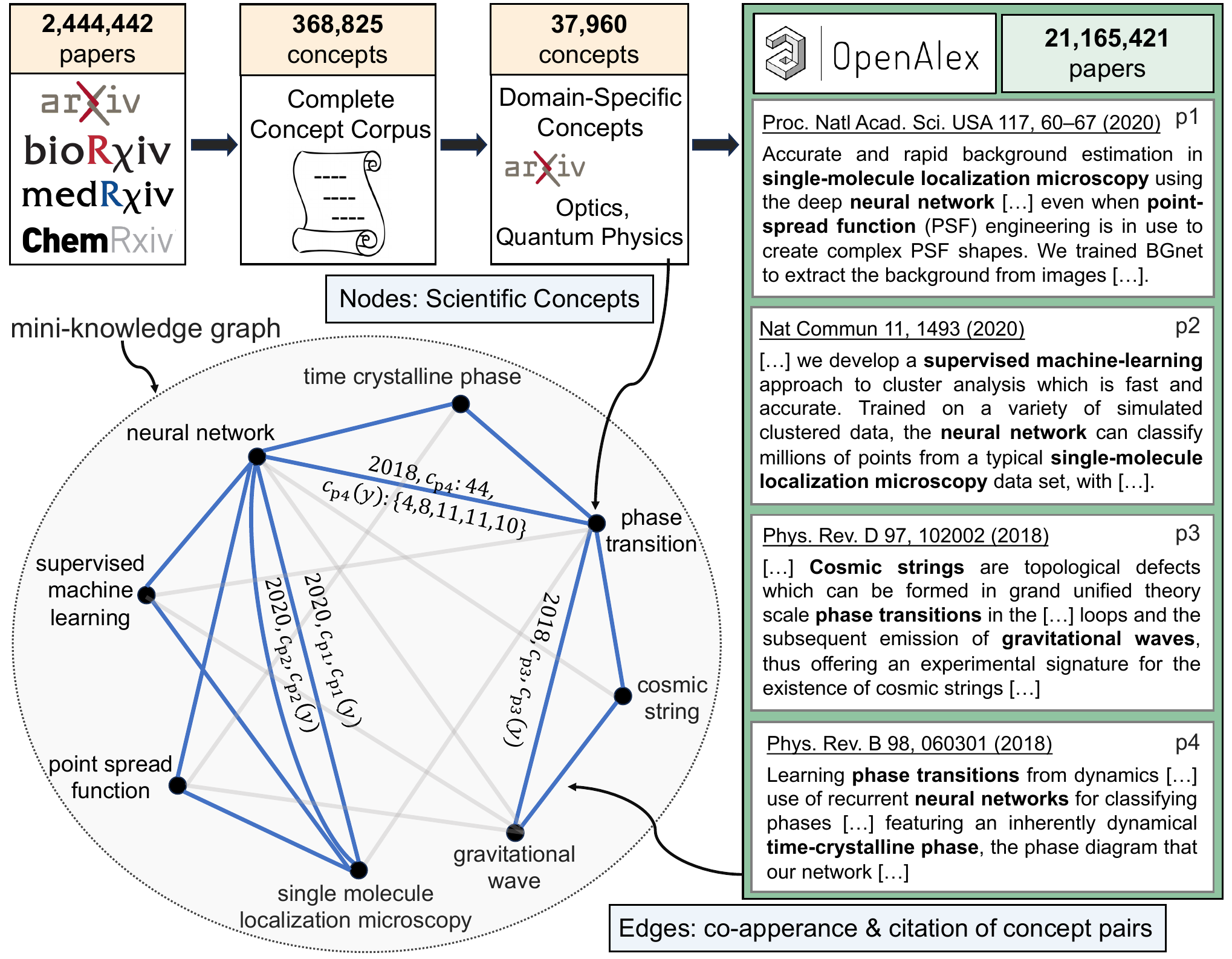}
    \caption{\textbf{Generation of the knowledge graph with time and citation information}. Vertices are formed by scientific concepts, which are extracted from scientific papers (titles and abstracts) from prominent academic preprint servers. Edges are formed when concepts are investigated jointly in a scientific publication. There are 21,165,421 out of 92,764,635 papers from OpenAlex which form at least one edge. The edges are augmented with citation information, which acts as a proxy for impact in our work. A mini-knowledge graph (blue edges) is constructed from four randomly selected papers (p1-p4) \cite{mockl2020accurate, williamson2020machine, abbott2018constraints, van2018learning} from OpenAlex as an example. Here, $c_{p4}$ represents the total citations of paper p4 since its publication, and $c_{p4}(y)$ is its annual citations from 2018 to 2022 (e.g., $c_{p4}(2018)=4$). The citation value of the edge is the sum of the all papers creating the edge.}
    \label{fig:semnet}
\end{figure*}
\begin{figure*}[!t]
    \centering
    \includegraphics[width=0.91\linewidth]{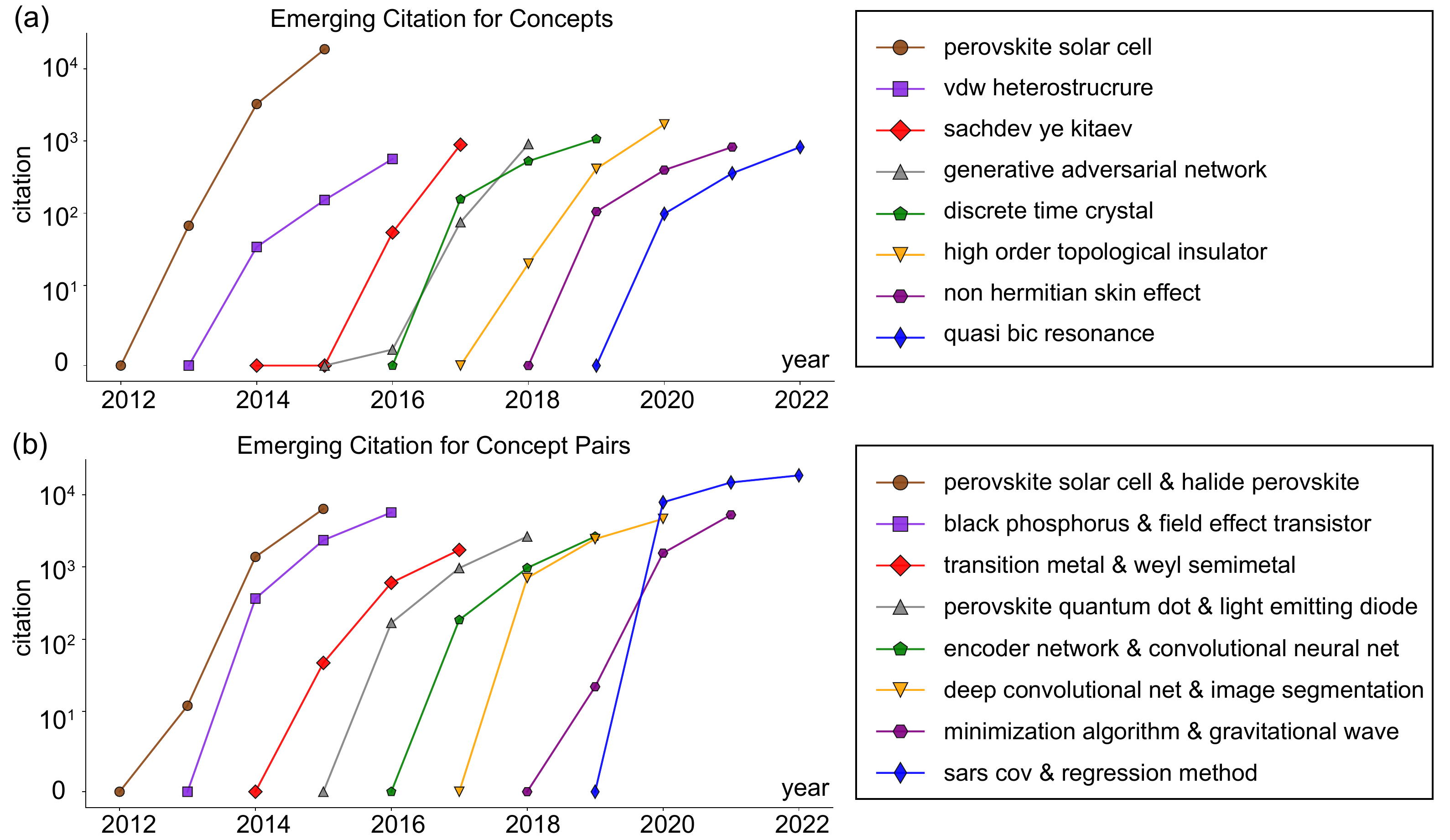}
    \caption{\textbf{Fastest growing citations of concepts and concept pairs}: Evolution of citations over three years for the top-fastest growing, previously uncited concepts (a) and concept pairs (b). We find many revolutionary topics in the realm of quantum physics and optics research in the last decade, including Perovskite devices \cite{rong2018challenges}, the emergence of complex and non-hermitian topology \cite{bergholtz2021exceptional}, the introduction of advanced concepts of machine learning in physics \cite{carleo2019machine,krenn2023artificial,wang2023scientific} and quasi-BIC (bound state in continuum) resonances. }
    \label{fig:data}
\end{figure*}

In this work, we answer this question by combining semantic networks and citation networks that are purely based on the level of scientific concepts\footnote{\href{https://github.com/artificial-scientist-lab/Impact4Cast}{GitHub: Impact4Cast}}\label{fn:github}. Specifically, we develop a large evolving knowledge graph using more than 21 million scientific papers, from 1709 (starting with a letter by Antoni van Leeuwenhoek \cite{leeuwenhoek1709ii}) to April 2023. The vertices of the knowledge graph are scientific concepts and the edges between two concepts contain information about when these topics have been investigated and how often they have been cited subsequently. We then train a machine learning model on the historic evolution of the knowledge graph. We find that the neural network can predict with high accuracy which concept pairs, that have never been jointly investigated before in any scientific paper, will be highly cited in the future. Being able to predict the potential impact of new research ideas -- before the paper is written or the research is done or even started -- could be a cornerstone in future scientific AI-assistants that help humans broadening their horizon of possible new research endeavours \cite{krenn2022scientific}.

\section*{Results}
\textbf{Creating a list of scientific concepts} --
At the heart of our knowledge graph are scientific concepts, as depicted in Fig.~\ref{fig:semnet}. We chose not to rely on existing concept lists, such as the APS or computer science ontology \cite{salatino2019cso}, for several reasons. Firstly, our goal is to ultimately cover all natural sciences comprehensively, and a universal list encompassing this breadth doesn't currently exist. Secondly, we want to capture the most recent concepts that might be absent from existing lists. Lastly, generating our list ensures that we have a granular understanding and control over the concepts.

To build our concept list, we started with 2,444,442 papers from four publicly available preprint servers: arXiv, bioRxiv, medRxiv, and chemRxiv. We use papers from preprint servers for two reasons: (1) It contains papers that are not published yet in journals, thus our dataset also contains state-of-the-art concepts; (2) they associate papers to research categories, which can be used to focus on specific scientific domains (as we do, with the field of quantum physics and optics). The data cutoff is February 2023. From these, we extracted titles and abstracts of the papers. To single out concept candidates from this extensive collection, we applied the Rapid Automatic Keyword Extraction (RAKE) algorithm based on statistical text analysis to automatically detect important keywords \cite{rose2010automatic} (see details in the Appendix~\ref{appB}). Each of these candidates are ranked Concepts with two words, like \textit{phase transition}, were retained if they appear in at least 9 papers, while longer concepts, such as \textit{single molecule localization microscopy}, needed to appear in at least 6 papers. In this way, we can increase the fraction of high-quality concepts. We further developed a suite of natural language processing tools to refine the concepts, followed by manual inspection to remove any incorrectly identified ones. Finally, we got a list which contains over 368,000 concepts. We focus here on concepts specific to the sub-field of optics and quantum physics (representing roughly 10\% of the entire concepts), but our method can immediately be translated to any other domain. This refined domain-specific concept list serves as the vertices of our knowledge graph.
\begin{figure*}[!t]
    \centering
    \includegraphics[width=0.8\linewidth]{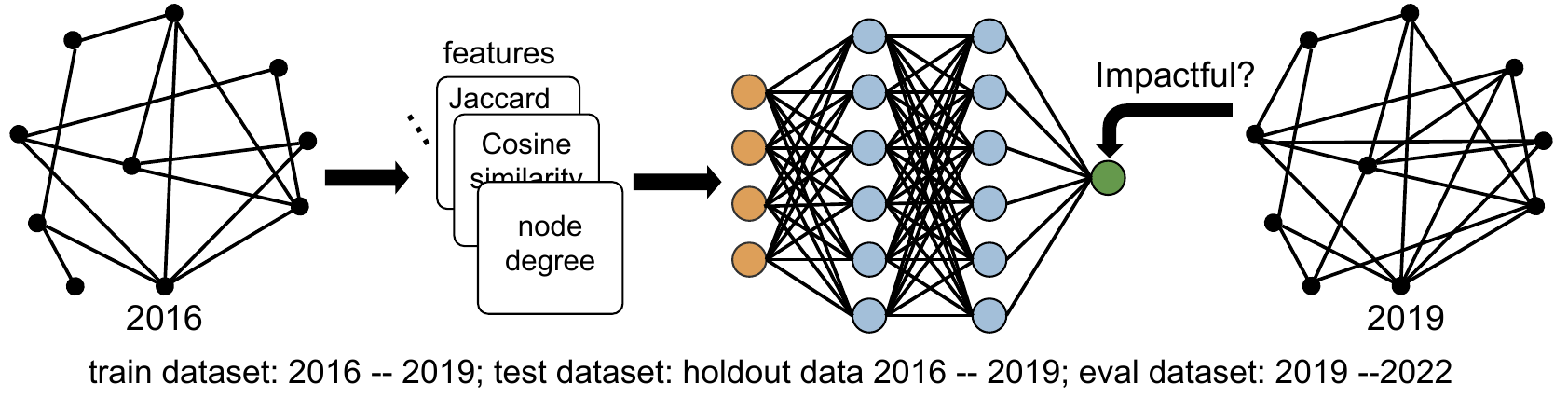}
    \caption{\textbf{Forecasting the impact of new research connections.} Network and citation features from unconnected vertex pairs from 2016 are used as input to a neural network. The citation information from 2019 is used as a supervision signal to train the neural network. After training, we evaluate the neural network's abilities by applying it to unconnected vertex pairs from 2019, aiming to predict the developments in 2022 -- a task involving data the network has never encountered before.}
    \label{fig:neuralnet}
\end{figure*}
\begin{figure*}[!t]
    \centering
    \includegraphics[width=0.8\linewidth]{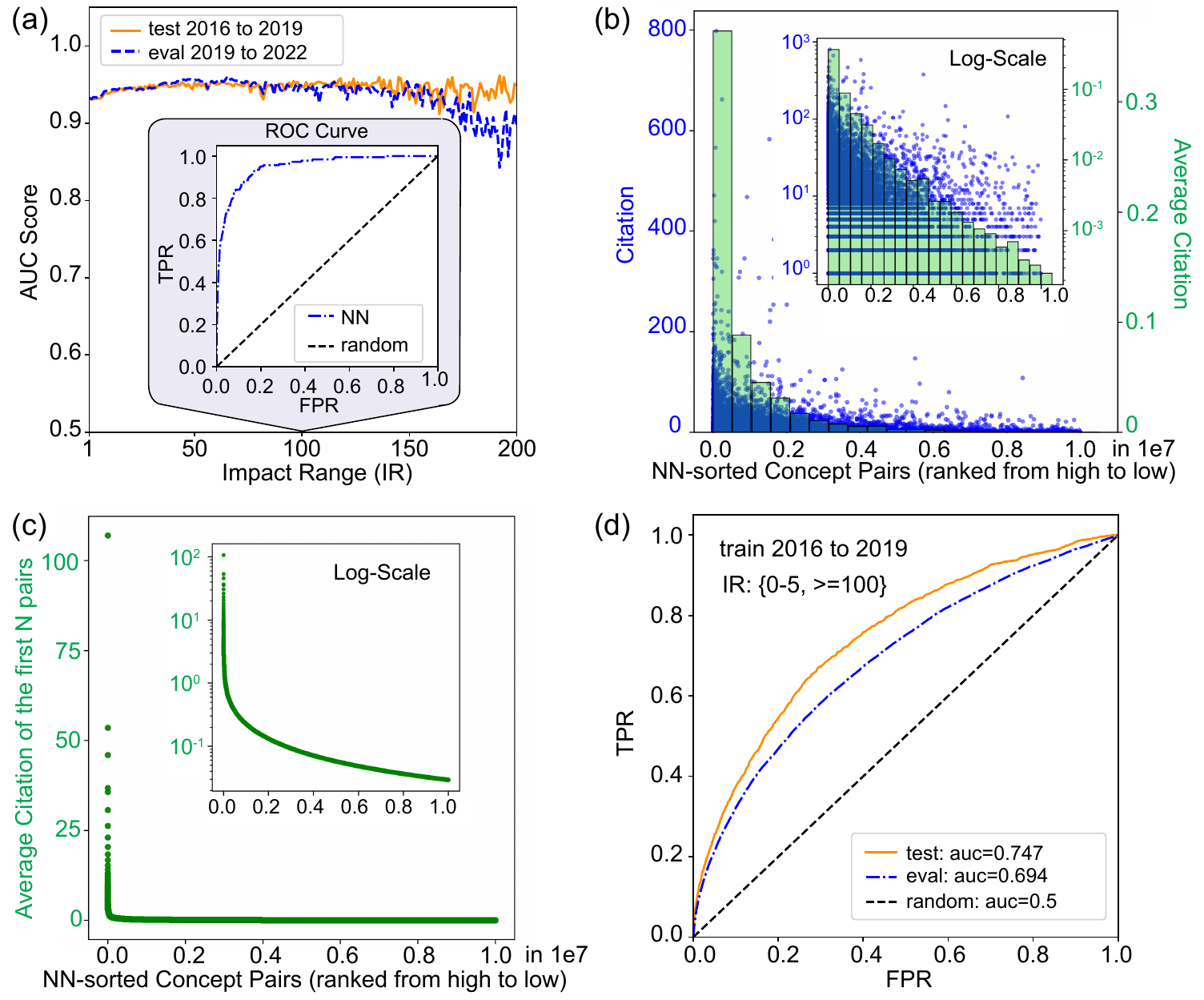}
    \caption{\textbf{Evaluating the machine-learning-based impact forecast.} (a): Classification of unconnected pairs, whether they will exceed a certain threshold three years later. Training data contains unconnected vertex pairs from 2016 and the supervision signal according to their impact range $IR$ 3 years later. The test dataset also includes pairs from 2016, but excludes those in the training set. A more challenging evaluation set contains unconnected pairs from 2019, with outcomes verified in 2022, importantly noting that the neural network was only trained on data from 2016 to 2019, not 2019 to 2022. We quantify the quality using the AUC of the ROC curve. For example, $IR=100$, i.e, ($<100$, $>=100$), refers to whether the 3-year citation counts after 2016 (test) or after 2019 (eval) is at least 100. TPR (true positive rate) measures how often a test correctly identifies a true positive, while FPR (false positive rate) measures how often it correctly identifies a true negative. (b): Sorted predictions of the neural network on the evaluation set (blue curve in (a)) shows the very high quality prediction at the level of individual concept pairs. The y-axis stands for the respective fraction of the evaluation dataset ($10^7$ data points). The histogram is separated into 20 equal bins. No fitting is involved. In (c), we show the average citation of the first N highest predicted concept pairs. This plot shows impressively that the highest predicted concept pairs indeed have very high citation, more than 3 orders of magnitude higher than the average citation of all $10^7$ pairs (0.029 citations). (d): This more challenging step shows that citation prediction goes beyond link predictions. Here we take unconnected vertex pairs, conditioned on a connection 3 years later. The neural network is tasked to classify these concept pairs in low or high citations, revealing that it is not just predicting new links, but is learning intrinsic citation features. Here $IR=[5,100]$, i.e, ($0-5$, $>=100$), means whether the 3-year citation count after 2016 (test) or after 2019 (eval) is at most 5 or at least 100.}
    \label{fig:auc_results}
\end{figure*}

\begin{figure*}[!t]
    \centering
    \includegraphics[width=0.90\linewidth]{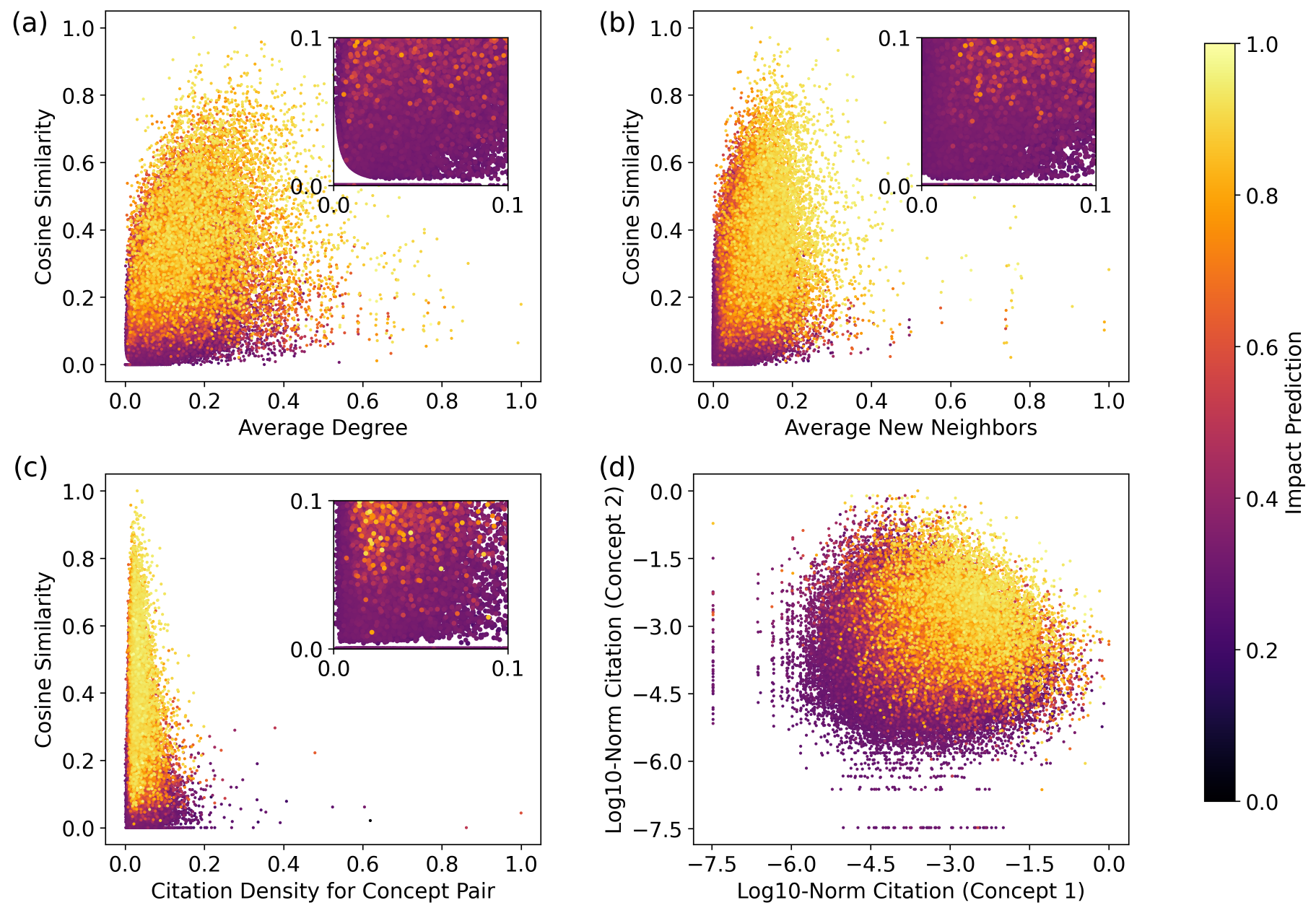}
    \caption{\textbf{Network features vs. predicted impact.} A randomly selected set of 100,000 unconnected pairs until January 2023 is used. The color represents the neural network's prediction of each concept pair's impact. (a): The y-axis shows cosine similarity, indicating the semantic similarity between concepts; lower values represent concept pairs that are semantically distinct. For each concept pair $(u,v)$, cosine similarity is calculated as the number of shared neighbors divided by the square root of the product of the number of neighbors of $u$ and $v$. The x-axis is the average vertex degree of the two concepts in the knowledge graph, reflecting their overall prominence. Concepts with low similarity and low degree yet predicted to have high impact could be surprising and offer interesting suggestions. (b): The x-axis represents the average number of new neighbors each concept gained over the last three years. Concept pairs with low similarity and few new neighbors but high impact predictions might highlight potentially overlooked but intriguing ideas. (c): The x-axis denotes \textit{citation density} (average citations per paper mentioning the concepts). Pairs with low similarity and citation density but high predicted impact could again indicate overlooked potential ideas. (d): Citation counts for concept 1 (x-axis) and concept 2 (y-axis) over last three years are plotted on a logarithmic scale. We can easily identify concept pairs predicted to have high impact in the future, even though they have individually received few citations in the past.}
    \label{fig:nn_results}
\end{figure*}

\begin{figure*}[!t]
    \centering
    \begin{minipage}{0.99\linewidth}
        \centering
        \includegraphics[width=1.0\linewidth]{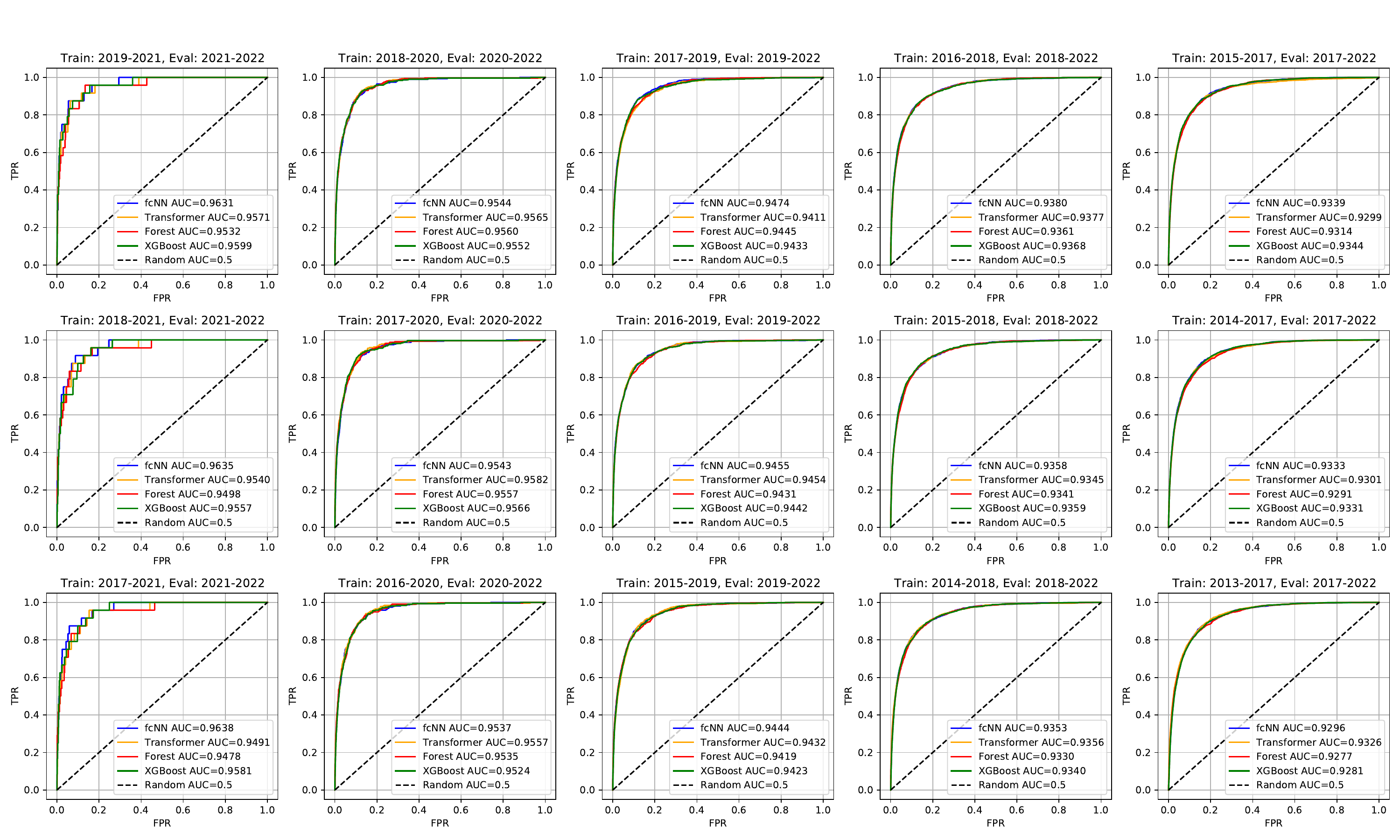}
        (a). Impact Range IR=10.
        \label{fig:combinedIR10}
    \end{minipage}
    \begin{minipage}{0.99\linewidth}
        \centering
        \includegraphics[width=1.0\linewidth]{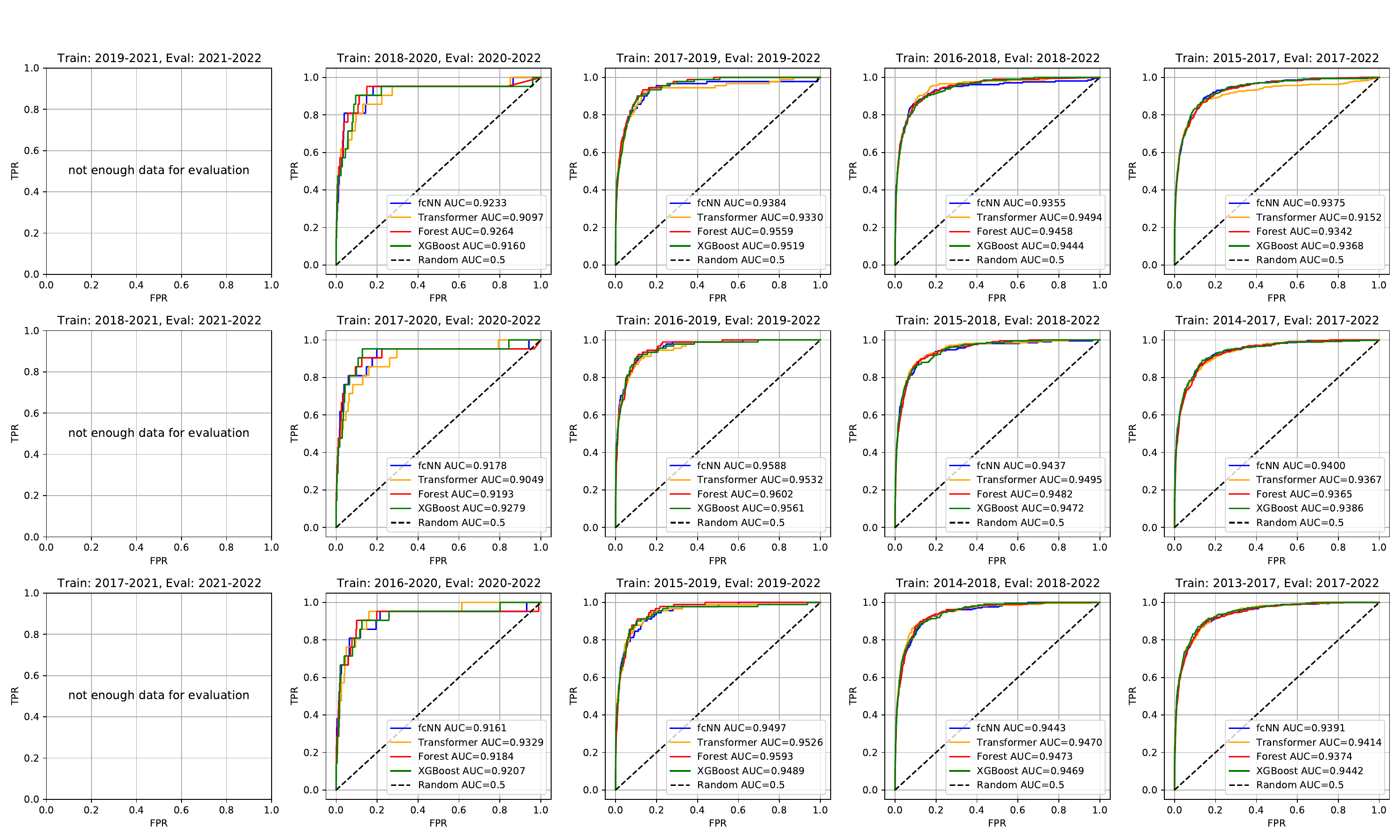}
        (b). Impact Range IR=50.
        \label{fig:combinedIR50}
    \end{minipage}
    \caption{\textbf{Benchmarking} fully connected NNs, Transformers, Random Forest, and XGBoost on 27 variations of the prediction task, with 2-4 year training and 1-5 year evaluation intervals, across two different impact ranges (IR).}
    \label{fig:combined_intervals}
\end{figure*}

\begin{figure*}[!t]
    \centering
    \includegraphics[width=1\linewidth]{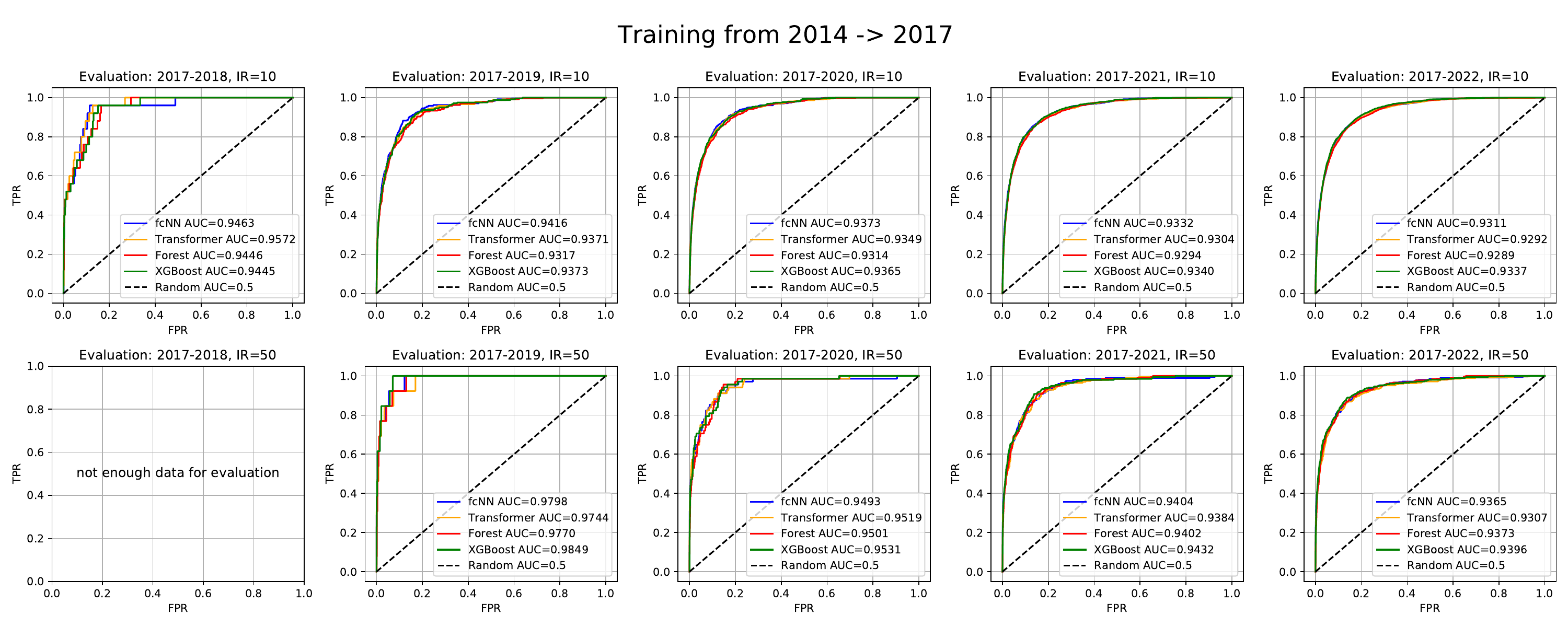}
    \caption{\textbf{Predictions for varying intervals without retraining}. The four ML models were trained using data from 2014 to 2017 and then used to predict outcomes 1 to 5 years into the future without retraining. For one task, the number of positive cases was insufficient for meaningful classification (we set a threshold of at least 10 positive cases out of 1 million total cases).}
    \label{fig:multi_train}
\end{figure*}
\begin{figure*}[!t]
    \centering
    \includegraphics[width=1\linewidth]{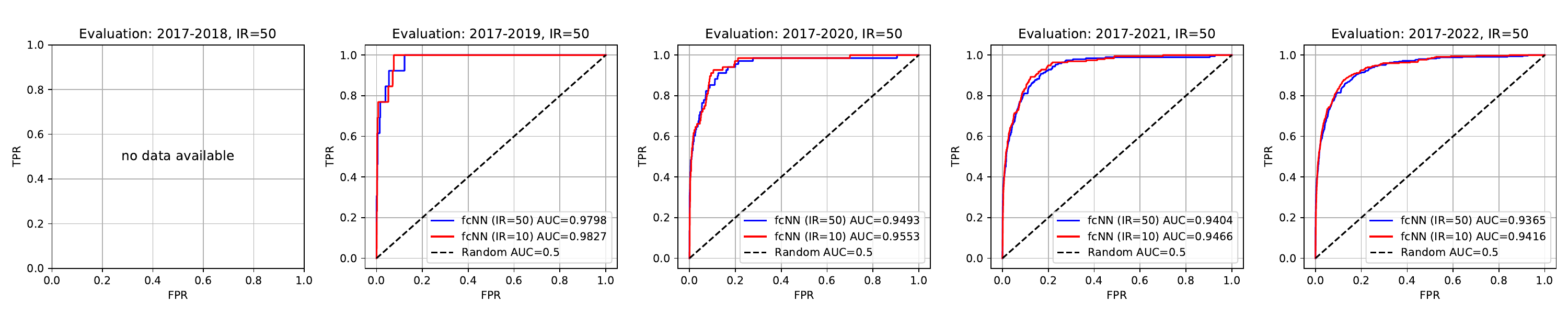}
    \caption{\textbf{Prediction of higher IR}. The models are tasked with predicting whether concept pairs will receive at least 50 citations. Training is performed on data from 2014 to 2017, with evaluation conducted over intervals of 2, 3, 4, or 5 years (a 1-year interval lacks sufficient data). The blue line represents the performance of a fully connected neural network trained specifically for this task. In contrast, the red line represents a fully connected neural network (with identical architecture and training parameters) trained to solve the task for $IR=10$ instead. Interestingly, the red model achieves slightly higher AUC values, indicating that predictions for higher impact ranges ($IR=50$) benefit from training on more diverse data, including those from lower impact ranges.
    }
    \label{fig:mixed_single}
\end{figure*}

\textbf{Creating an evolving, citation-augmented knowledge graph} --
Now that we have the vertices, we can create edges that contain information from the scientific literature. We get the citation information from papers in OpenAlex \cite{openalex}, an open-source database containing detailed information on more than 92 million publications. Edges are drawn when two concepts co-occur in the title or abstract of a scientific paper. If a paper connects two vertices, the weight of the newly formed edge is the paper's annual citation numbers from 2012 to 2023 together with the total citation number since its publication. If more than one paper creates an edge, then the edge contains the sum of the annual citations (as well as the sum of the total citations) gained by all papers. As research papers appear over time, and their citations are created in time, we effectively build an evolving, citation-augmented knowledge graph that evolves in time (see Fig.~\ref{fig:semnet}). From these 92 million papers, 21 million contain at least two concepts of our concept list and can therefore for an edge in the knowledge graph.

The final constructed knowledge graph has 37,960 vertices with more than 26 million edges (built from 190 million concept pairs, containing multi-edges when multiple papers create the same edge) from the OpenAlex dataset, with a data cutoff at April 2023. 

In Fig.~\ref{fig:data}, we show the fastest growing (in terms of citation) concepts and concept pairs since 2012, where we can recognize many highly influential topics in quantum physics and optics research.

\textbf{Forecasting impact of newly created concept connections} -- 
With an evolving knowledge graph from the past, we can formulate the prediction of impact for new concept pairs as a supervised learning task, as illustrated in Fig. \ref{fig:neuralnet}. For a vertex pair that has not had any connection in the year 2016, we predict whether three years later this pair accumulated more than a certain number of citations. Using the historical knowledge graph, we possess an ideal supervision signal for our binary classification task. During the training phase, we selected pairs of vertices that were not connected and calculated 141 features for each pair. These features include 41 network features, divided into 20 node features (such as the number of neighbors and PageRank \cite{page1998pagerank} over the past three years) and 21 edge features (including cosine, geometric, and Simpson similarities \cite{barabasi2016network}). Additionally, we incorporated 100 impact features: 58 of these are node citation features, covering total citations and yearly citations within the last three years. The other 42 features are about vertex pairs and include measures such as the citation ratio between them. Detailed feature description are available in \href{https://github.com/artificial-scientist-lab/Impact4Cast}{GitHub: Impact4Cast} and Tables.~\ref{tab:features} and ~\ref{tab:features2} in Appendix~\ref{appD}. The network features are inspired by the winner of the \textit{Science4Cast competition} \cite{lu2021predicting, krenn2020predicting}, and the citation features are developed empirically and could potentially be improved by careful feature importance analysis. Our neural network is a fully connected feed-forward network with four hidden layers of 600 neurons each. The exploration of more advanced architectures might improve the prediction qualities further. The neural network has to predict whether the unconnected vertex pair in 2019 will have at least $IR$ citations ($IR$ stands for the impact range).

The \textbf{impact range (IR)} is a threshold representing the minimum number of citations a concept pair must accumulate within a specified time frame (e.g., three years) to be classified as ``high-impact". For instance, an $IR=100$ means that only concept pairs with at least 100 citations during the defined period are considered impactful. This binary threshold simplifies the problem into a classification task, making it computationally tractable while providing a clear measure of success. Predicting individual citation counts is inherently noisy due to the stochastic nature of citation dynamics. By using IR, we focus on identifying high-impact trends, avoiding fluctuations of precise citation counts. IR provides a clear and measurable target for classification. This allows us to use metrics like the Area Under the Curve (AUC) of Receiver Operating Characteristics (ROC) curves to evaluate the prediction quality \cite{fawcett2004roc}.

We perform the training for different values of the impact range IR from $IR=1$ to $IR=200$, and then quantify the quality with AUC of the ROC curves \cite{fawcett2004roc}. The AUC gives a measure of classification quality and stands for the probability that a randomly chosen true example is ranked higher than a randomly chosen false example. A random classifier has $AUC=0.5$. We measure the AUC for a test set (which contains unconnected pairs not in the training set) for a prediction from 2016 to 2019, and for an evaluation dataset, with 10 million random data from 2019 to 2022 (while keeping the training data of the neural network from 2016 to 2019). The evaluation dataset shows how well the neural network performs on future, never-seen datasets. This is motivated by our goal that ultimately we want to train a neural network with all available data (let's say, until January 2023) and predict what happens until the future in 2026. In Fig.~\ref{fig:auc_results}(a), we find that the AUC scores for both the test set and the evaluation set are beyond 0.8, in most of the cases beyond 0.9, for different $IR$. We can conclude that the neural network can forecast a high impact of previously never-investigated concept connections to a high degree. In Fig.~\ref{fig:auc_results}(b), we sort the concept pairs of the evaluation dataset with the neural network ($IR=100$), and plot their true citation counts. We further divide the 10 million evaluation dataset into 20 equal parts and plot their average citation count (represented by green bars) for each 5\% segment. This clearly demonstrates good predictions at the individual concept pair level. As seen in Fig.~\ref{fig:auc_results}(c), the highest predicted concept pairs indeed get more than 3 orders of magnitude more citations than the average citation of all 10 million pairs.

\textbf{Forecasting genuine impact beyond link prediction} -- 
Next, we perform an even more challenging, genuine impact prediction task that goes beyond link prediction (i.e., predicting which concept pairs will be investigated in the future by a scientific paper). Concretely, in this task training data is conditioned on unconnected vertex pairs in 2016 which are actually connected in 2019. The neural network only gets citation information from 2016 and has to predict whether the newly generated concept pair will be highly impactful or not in the future. For that, our classification task asks whether the newly generated edge will receive citations within 0-5 or above 100 (Fig.~\ref{fig:auc_results}(d)) in 2019. We see that the AUC score is beyond 0.7 (for the test set) and beyond 0.67 for the evaluation set, clearly indicating that the neural network can predict impact properties that go beyond the simple link-prediction task.

\textbf{Highly predicted impact pair and potential applications} -- 
We can now investigate the largest predicted pairs of concepts, by taking all unconnected vertex pairs ($\sim$694 million pairs) until January 2023, and let the neural network (trained with all unconnected pairs in 2019 with supervision signal in 2022) sort them by impact predictions. We find that the highest predicted pair is \textit{renewable energy} and \textit{cancer cell}. This prediction is a very high-risk bet. For more practical, personalized suggestions, one can restrict the unconnected concept pairs to those related to specific scientists or research groups, aiming for high-impact collaboration suggestions. By examining the published works of scientists to identify their research interests, it becomes possible to identify concept pairs where one aligns with one scientist's specialty and the other with another scientist's. Thereby, one can suggest potential collaborations of high impact. As an example, by constraining the personalized research interests of scientists in experimental quantum optics and one researcher in biophysics, the highest predicted impact concepts pairs are `\textit{microfluidic channel}' with `\textit{Kerr resonator}', \textit{`SARS CoV'} with `\textit{quantum enhanced sensitivity}' or `\textit{electron microscopy}' with `\textit{quantum vacuum field}'. These suggestions can be further refined based on their similarity (e.g., represented by the cosine similarity) or the prominence of the concepts (indicated by the node degree), as we show in Fig.~\ref{fig:nn_results}. Here, we plot 100,000 concept pairs that have not been studied together until January 2023 and use the neural network trained on 2019 dataset to predict their impact. The points are plotted based on various properties, such as the similarity between concepts, their prominence within the network, their growth rate in the network (reflected in newly acquired neighbors), and how often the concepts have been cited previously. Plotting in this way allows us to identify rare outliers -- concept pairs with high predicted impact that have unique properties, such as the bright yellow spots highlighted in the insets of Fig.~\ref{fig:nn_results}. These methods help us narrow down the enormously large number of possibilities into a small number of personalized and targeted suggestions, which could inspire new ideas. In practical application, it will be useful to update the knowledge graph regularly, and train the machine learning models on the latest knowledge graph data, so it can better incorporate the latest trends and discoveries for its predictions.

\textbf{Benchmarking different models and different time intervals} --
So far, we have only focused on a specific case: a training interval from 2016 to 2019 (3 years) and an evaluation interval from 2019 to 2022 (3 years). Additionally, we have primarily investigated the performance of feed-forward neural networks. Exploring other models and examining their predictive capabilities across various training and evaluation intervals could provide deeper insights into model performance. To do so, we expanded our study to include benchmarking on a small dataset of 1 million pairs. This benchmarking incorporates the previously used fully connected neural network alongside additional models, including a transformer architecture \cite{bahdanau2016neuralmachinetranslationjointly,NIPS2017_3f5ee243}, random forest \cite{quinlan1986induction}, and XGBoost \cite{chen2016xgboost}.

The feed-forward neural network, implemented using PyTorch \cite{NEURIPS2019_9015}, consisted of three hidden layers, each with 600 neurons and ReLU activations \cite{nair2010rectified}, resulting in approximately 800,000 trainable parameters. Similarly, the transformer architecture \cite{bahdanau2016neuralmachinetranslationjointly,NIPS2017_3f5ee243} also implemented via PyTorch, was designed with 4 layers, a hidden size of 128, 4 attention heads, and a feedforward dimension of 512, resulting in approximately 800,000 trainable parameters. Positional encodings were added to the input features. Both neural network models were trained using Adam optimizer \cite{kingma2014adam} with a batch size of 2048 and a learning rate of 0.0001. The random forest classifier, implemented with scikit-learn \cite{pedregosa2011scikit}, was trained with 300 trees, a minimum of 25 samples required to split a node, and 10 samples per leaf. The XGBoost model was trained using up to 2000 boosting rounds, a learning rate of 0.01, and a maximum tree depth of 10. Hyperparameters for all models were selected via a hyperparameter search for a single benchmark task (training: 2016–2019, evaluation: 2019–2022, with $IR=10$) and kept constant for all tasks.

In all tasks, the models are provided with 141 input features of a specific concept pair and have to predict whether this pair will receive more or fewer IR citations ($IR=10$ or $IR=50$) in certain future years. To achieve this, the four models are trained on 2-, 3-, and 4-year intervals and evaluated on intervals ranging from 1 to 5 years. For example, if the training interval spans 2 years and the evaluation interval spans 5 years, the models are trained using data from 2015 to 2017 to predict whether unconnected concept pairs will receive IR citations. They are then evaluated using 2017 data to make predictions for 2022. After training, the models are evaluated on 1 million concept pairs, predicting the likelihood of each pair receiving more than IR citations. These predictions are ranked (from high to low likelihood) and compared against the ground truth to compute the ROC curve and the AUC score, which measures prediction quality. As shown in Fig.\ref{fig:combined_intervals}, the models achieve AUC values exceeding 90\% in these tasks. In a slightly modified task, we train the models on the data from 2014 to 2017 and evaluate them from 2017 to 1-5 years into the future (2018 to 2022). This test analyzes how well the prediction perform for intervals on which the models have not been trained and how difficult it is to predict further into the future. As shown in Fig.\ref{fig:multi_train}, the quality of the predictions indeed decreases for larger intervals.

In Fig.\ref{fig:mixed_single}, we show that models trained on smaller impact ranges can predict higher impact ranges even slightly better than those trained exclusively on higher impact ranges. This might be due to more diverse training data (there are many more examples of $IR=10$ than of $IR=50$, because many more concept pairs achieve at least 10 citations rather than 50). It might also be explained by a systematic drift in citation patterns between the years the models were trained and the years they are applied, potentially due to the growing number of overall citations. It will be very interesting in the future to explore and understand this effect further and to develop new ML methods that could leverage this dynamic.

All of our models use the same set of input features and do not directly access the full knowledge graph. Developing techniques that can leverage more general graph properties -- such as automatically learning features or generating embeddings -- would be an interesting avenue to explore. A related approach was demonstrated in a previous competition \cite{krenn2023forecasting}, where the task was to predict the future state of a knowledge graph. There, the graph’s edges represented co-occurrences of concepts in scientific papers. In contrast, the knowledge graph used in our study is more complex, with edges also weighted by the number of citations received by concept pairs. Exploring more end-to-end approaches that integrate more information from the entire knowledge graph could reveal whether such methods can outperform the models and hand-crafted features demonstrated in this work.

\section*{Discussion}
We show how to forecast the impact of future research topics. Although we view this as a significant step towards developing truly useful AI-driven assistants, achieving this goal requires numerous further advancements. Firstly, developing methods to extract more complex information from each paper will be crucial, for instance by employing hyper-graph structures that carry more information from each paper \cite{battiston2021physics, belikov2022prediction}, which has already been demonstrated to lead to exciting results in other domains \cite{foster2015tradition, wang2021science, sourati2023accelerating}. The forecast itself could benefit from more genuine dynamical features that go beyond network snapshots from different years \cite{nguyen2018continuous}, or the application of dynamic word embedding \cite{frohnert2025discovering}. This might also allow for the forecast of new concepts \cite{salatino2017topics, salatino2018augur} and their impact. Incorporating the recent dataset \cite{lin2023sciscinet, li2019dataset} into our research could also allow us to explore more complex data structures than those used in our paper. Secondly, it will be interesting to approximate impact with metrics that go beyond citations -- which is a crucial topic in computational sociology and the study of the science of science \cite{fortunato2018science, wang2021science}. Additionally, introducing metrics of surprise, as discussed in \cite{foster2021surprise, shi2023surprising}, could serve as a complementary metric to citation prediction for ranking suggestions. Finally, while the suggestion of \textit{impactful} new ideas might be a key component of future AI assistants, it will be crucial to study its relation to the \textit{scientific interest} of working researchers \cite{gu2024generation}.

\section*{Acknowledgements}
The authors thank Burak Gurlek for interesting discussions at the start of this project, and the organizers of OpenAlex, arXiv, bioRxiv, chemRxiv, and medRxiv for making scientific resources freely and readily accessible. X.G acknowledges the support from the Alexander von Humboldt Foundation.

\textbf{Author Contributions} X.G. and M.K. designed research; X.G. performed research and analyzed data; and X.G. and M.K. wrote the manuscript.\\
\textbf{Competing Interests} The authors declare that they have no competing financial interests.

\textbf{Data availability statement} Data is accessible on Zenodo at \href{https://doi.org/10.5281/zenodo.10692137}{https://doi.org/10.5281/zenodo.10692137} \cite{Impact4Cast_Zenodo}. Benchmark data used in our work is also available on Zenodo at \href{https://doi.org/10.5281/zenodo.14527306}{https://doi.org/10.5281/zenodo.14527306} \cite{Impact4Cast_benchmarkZenodo}. Codes for this work are available on GitHub at \href{https://github.com/artificial-scientist-lab/Impact4Cast}{https://github.com/artificial-scientist-lab/Impact4Cast}.

\appendix
\section{Datasets for the knowledge graph}
\label{appA}
To compile a list of scientific concepts in natural science, we used metadata from four major preprint servers: arXiv, bioRxiv, medRxiv, and chemRxiv. The arXiv dataset can be directly downloaded from \href{https://www.kaggle.com/datasets/Cornell-University/arxiv}{Kaggle}, while metadata from bioRxiv, medRxiv, and chemRxiv are accessible through their APIs. The full methodology and codes are available on the \href{https://github.com/artificial-scientist-lab/Impact4Cast}{GitHub: Impact4Cast}. Our comprehensive dataset encompasses approximately 2.44 million papers, including 78,084 from arXiv's physics.optics and quant-ph categories, which were specifically utilized for identifying domain concepts. 

For edge generation, we used the OpenAlex database snapshot, available for download in \href{https://openalex.s3.amazonaws.com/browse.html}{OpenAlex bucket}. More details can be found at the \href{https://docs.openalex.org/download-all-data/snapshot-data-format}{OpenAlex documentation site}. The complete dataset occupies around 330 GB, expanding to approximately 1.6 TB when decompressed. Our interest was specifically in scientific journal papers that include publication time, title, abstract, and citation information. By focusing on these criteria, we managed to reduce the dataset to a more manageable gzip-compressed size of 68 GB, comprising around 92 million scientific papers. From these 92 million papers, 21 million contain at least two concepts of our final concept list and can therefore for an edge in the knowledge graph.

\section{Details on concept and edge generation} \label{appB} 
From the preprint dataset of $\sim$2.44 million papers, we analyzed each article's title and abstract using the RAKE algorithm.

RAKE works by proposing candidate key phrases from each sentence by splitting the sentence at punctuation marks and stop words. Let's take the sentence: \textit{Recurrent neural networks have significantly improved the accuracy of image recognition in large-scale scientific collaboration networks.} RAKE splits this into the candidate phrases \textit{Recurrent neural networks}, \textit{improved}, \textit{accuracy}, \textit{image recognition}, and \textit{large-scale scientific collaboration networks}. Each individual word within these candidate phrases receives a score based on two metrics: its frequency (how often it occurs in candidate phrases; here only \textit{networks} occurs twice, all others once) and its degree (the number of co-occurrences with other words in candidate phrases). For example, the word \textit{networks} has a higher degree because it appears in multiple longer phrases, thus increasing its importance. The final keyword phrases are ranked based on the summed scores of their individual words, naturally favoring longer, meaningful phrases such as \textit{large-scale scientific collaboration networks} or \textit{recurrent neural networks}.

RAKE's candidates are then used for subsequent analysis. We filtered out concepts to retain only those with two words that appeared in nine or more articles, and those with three or more words that appeared in six or more articles. This step significantly reduced the noise from the RAKE-generated concepts, yielding a refined list of 726,439 relevant concepts. To further enhance the quality of the identified concepts, we developed a suite of automatic tools designed to identify and eliminate common, domain-independent errors often associated with RAKE. In addition, we conduct a manual review to identify and eliminate any inaccuracies in the concepts. The entire process, which included eliminating non-conceptual phrases, verbs, ordinal numbers, conjunctions, and adverbials, resulted in a full list of 368,825 concepts. 

We note that our approach might result in similar or synonymous concepts appearing as separate key phrases due to slight variations or phrasing differences. Although multiple entries in the extracted concept list may represent essentially the same idea, synonymity can effectively be identified by leveraging structural information within the knowledge graph. Specifically, network cosine similarity between nodes (concepts) reveals the degree of similarity in their network neighborhoods, indicating semantic closeness. Practically, this can be utilized to enhance impact forecasting: when recommending new high-impact concept pairs, applying a cosine similarity threshold ensures that suggested pairs have sufficiently distinct meanings, thus avoiding redundancy from synonymous phrases.

We then specifically focused on articles within the physics.optics and quant-ph categories from arXiv to extract domain-specific concepts. Iterating this entire list of concepts to these domain-specific articles, we identified 87,741 relevant concepts. Employing our specially designed automated filtering tool for initial refinement and then conducting a thorough manual review to remove inaccuracies, we narrowed the list down to 37,960 high-quality, domain-specific concepts.

As an example, we show the extraction of concepts for the four papers used in Fig.~\ref{fig:semnet}:
\begin{enumerate}
\item Accurate and rapid background estimation in single-molecule localization microscopy using the deep neural network BGnet \cite{mockl2020accurate}: `super resolution reconstruction', `neural network', `single molecule tracking', `deep neural net', `deep neural network', `localization microscopy', `biological structure', `point source', `point spread function', `single molecule localization microscopy', `optical microscopy', and `neural net'.
\item Machine learning for cluster analysis of localization microscopy data \cite{williamson2020machine}: `neural network', `supervised machine learning', `spatial relation', `localization microscopy', `single molecule localization microscopy', `neural net', and `machine learning'.
\item Constraints on cosmic strings using data from the first Advanced LIGO observing run \cite{abbott2018constraints}: `phase transition', `cosmic string', `gravitational wave', `cosmic microwave', `cosmic microwave background', `topological defect', `string theory', and `ring theory'.
\item Learning phase transitions from dynamics \cite{van2018learning}: `neural network', `recurrent network', `time crystalline phase', `phase transition', `localization transition', `spin chain', `recurrent neural net', `ct model', `recurrent neural network', `phase diagram', `crystalline phase', `neural net', and `recurrent net'.
\end{enumerate}

We created concept pairs, or edges, from the OpenAlex dataset, by detecting when domain-specific concepts co-occurred in paper titles or abstracts. This yielded 193,977,096 concept pairs (including multi-edges) across about 21 million papers. Each edge receives a time-stamp based on its paper's publication date, converted to the number of days since January 1, 1990. The final full knowledge graph comprises 26,010,946 unique edges after merging multiple edges between the same concept pairs. The citation information for an edge includes the paper's yearly citations from 2012 to 2023, alongside its total citation since publication. The OpenAlex dataset excludes yearly citations older than ten years, hence the focus on this specific ten-year time frame due to the absence of data prior to 2012. For edges formed by multiple papers, the edge weight combines the annual and total citations from all contributing papers.

Consider the edge formed by the concepts `single molecule localization microscopy' and `neural net', generated from paper $p1$ \cite{mockl2020accurate} published on January 7, 2020. The time-stamp for this edge is derived from the days elapsed since January 1, 1990. The citation metrics for this edge includes the total and yearly citations from each contributing paper. Paper $p1$ with 38 citations ($c_{p1}$=38), contributes yearly citations represented as $c_{p1}(y_i)$ for $i$=2023,2022,..., 2012, with actual values $\{5, 8, 16, 9\}$ for 2023 to 2020, and zeros for previous years, culminating in a citation sequence $\{5, 8, 16, 9, 0, 0, 0, 0, 0, 0, 0, 0\}$. Similarly, paper $p2$ \cite{williamson2020machine}, published on March 20, 2020, with 43 citations ($c_{p2}$=43), adds its yearly citations $\{6, 16, 14, 7\}$ for the same period (2023 to 2020). The aggregated citation data for this edge, combining $c_{p1}$ and $c_{p2}$, yields a total of 81 citations, with an annual citation sequence of $\{11, 24, 30, 16, 0, 0, 0, 0, 0, 0, 0, 0\}$.

\begin{figure}[!t]
    \centering
    \includegraphics[width=1.0\linewidth]{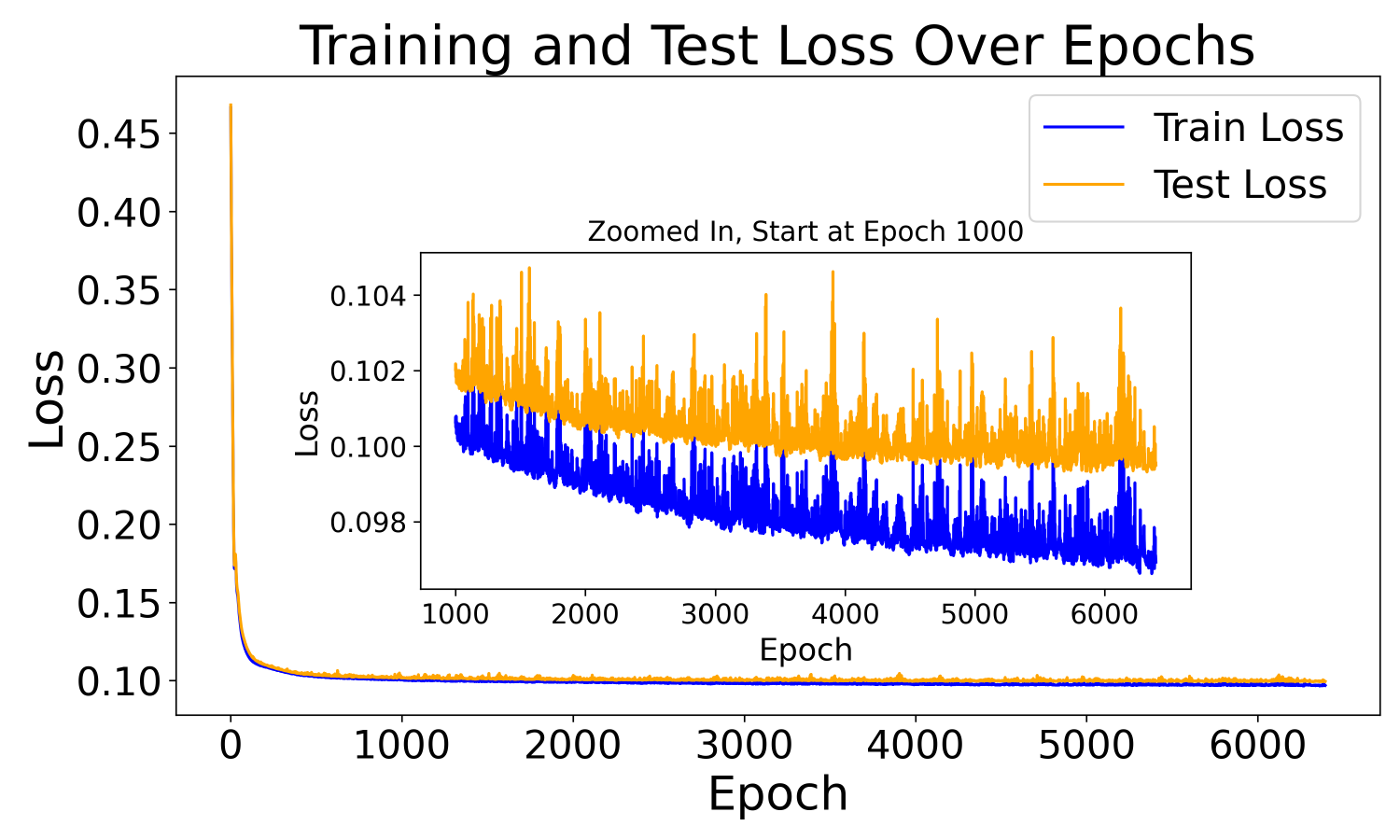}
    \caption{\textbf{Loss curves for one typical example.} The training and test loss curves correspond to a fully connected neural network trained on data from 2017 to 2020, with an impact range (IR) of 10.} 
    \label{fig:traintestloss}
\end{figure}
\begin{figure}[!t]
    \centering
    \includegraphics[width=1.0\linewidth]{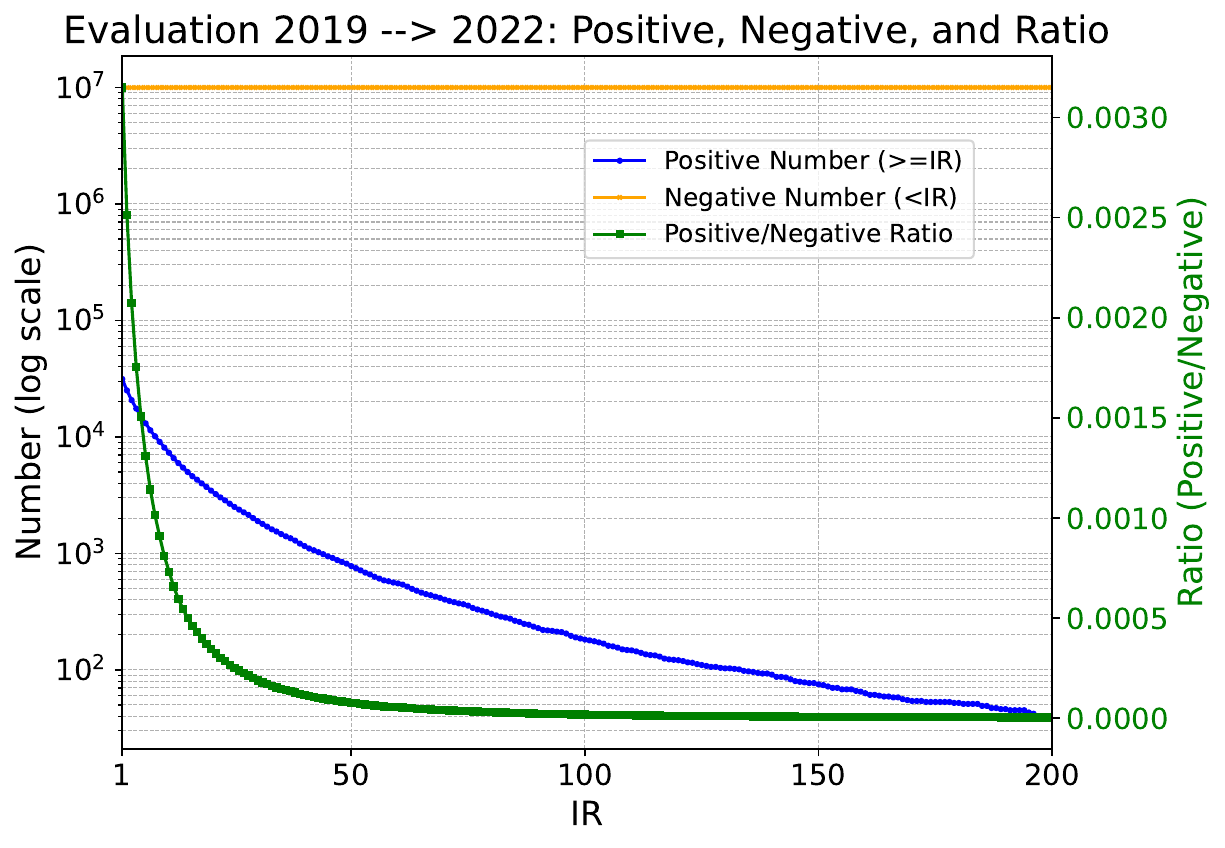}
    \caption{\textbf{Positive, negative samples, and their ratio in the 10M evaluation dataset (2019-2022) versus IR.}} 
    \label{fig:posnegratio}
\end{figure}
\begin{figure*}[!t]
    \centering
    \includegraphics[width=0.88\linewidth]{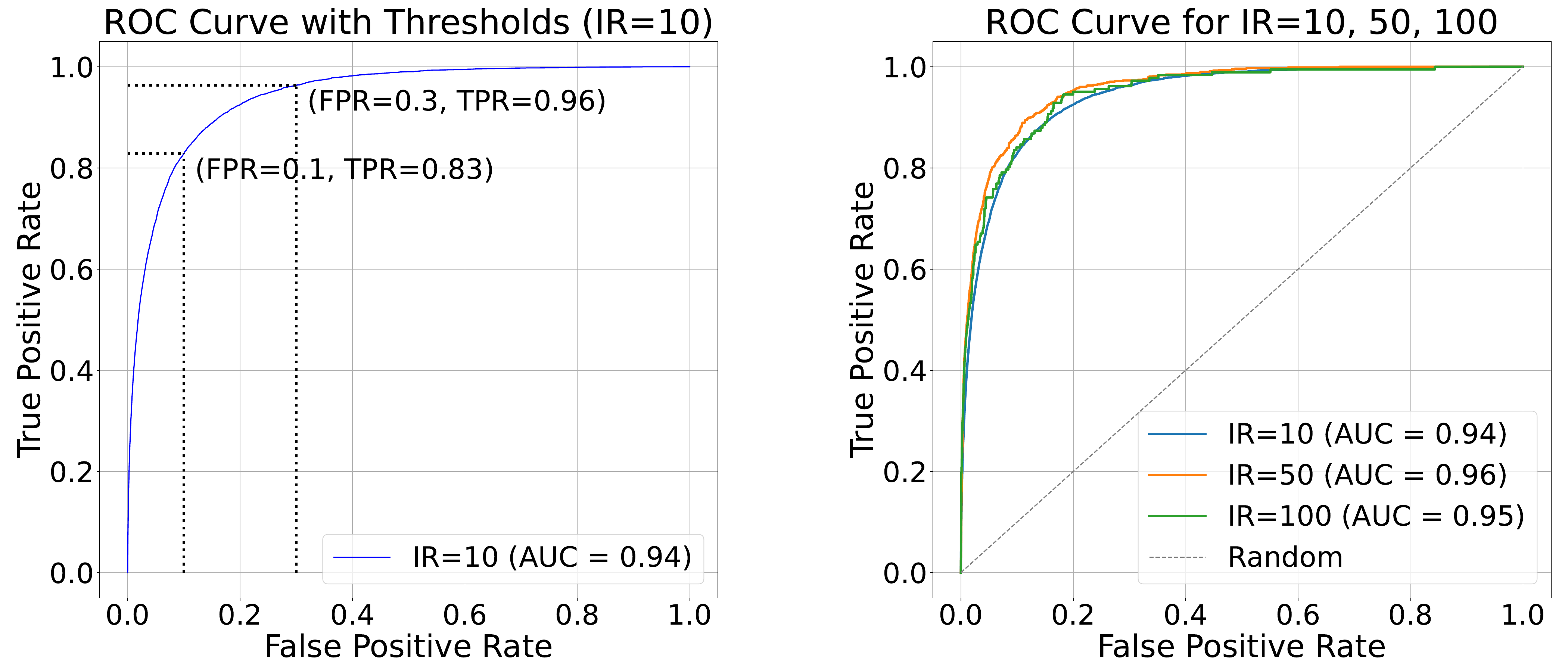}
    \caption{\textbf{ROC curve explanation.} The ROC curve are plotted for the evaluation data in Fig.\ref{fig:auc_results}, which illustrates the performance of binary classifiers discussed in our paper. On the left, for $IR=10$, the area under the curve (AUC) is 0.94.  The x-axis represents the False Positive Rate (FPR), while the y-axis represents the True Positive Rate (TPR). The curve demonstrates how the classifier’s performance changes with variations in the classification threshold, which determines whether a case is classified as Class 0 or Class 1. For instance, at a threshold that gives a classifier with FPR=0.1, the TPR is 0.83, and at FPR=0.3, the TPR is 0.96. Therefore, the ROC curve is more informative than just a single pair of FPR and TPR. On the right, the ROC curves for IR=10, 50, and 100 are shown.} 
    \label{fig:rocExplain}
\end{figure*}
\section{Details on training} \label{appC} 
Our neural network consists of six fully connected layers, which include four hidden layers with 600 neurons each. The network inputs are 141 features for each unconnected concept pair $(v_i,v_j)$, denoted as $p_{i,j}=(p^{1}_{i,j}, p^{2}_{i,j}, ..., p^{141}_{i,j})$, where each $p_{i,j}\in \mathbb{R}$. For instance, $p^{1}_{i,j}$ and $p^{2}_{i,j}$ represent the vertex degree of concepts $v_i$ and $v_j$ for the current year, $y$. Detailed feature description and feature generation code are available in \href{https://github.com/artificial-scientist-lab/Impact4Cast}{GitHub: Impact4Cast}. 

In our training process for the year 2016 to predict impact in 2019, we prepared a dataset comprising approximately 689 million unconnected concept pairs. The goal was to evaluate these pairs to determine whether their 3-year citation counts would have at least $IR$ citations ($IR$ is impact range) or not. From this extensive collection, we selected all positive samples (the 3-year citation counts are at least $IR$). An equivalent number of negative samples were then randomly chosen to match the size of the positive set. The refined dataset was subsequently divided, allocating 85\% for training and 15\% for testing purposes. For the evaluation dataset in 2019, which aims to predict the impact in 2022, we randomly selected 10 million unconnected pairs. Our neural network was trained using the Adam optimizer with a learning rate of $3\times10^{-5}$ and a mini-batch size of 1000. In every training batch, we randomly chose an equal number of positive and negative samples from the training set. This approach was also applied to our 2019 training process for predictions into 2022, where the trained neural network is used for future forecasting. An example loss curve is shown in Fig.\ref{fig:traintestloss}. An example for the number of positive and negative cases of the evaluation dataset (for the experiments in Fig.\ref{fig:auc_results}) is shown in Fig.\ref{fig:posnegratio}. In Fig.\ref{fig:rocExplain} we explain more details of the ROC curve, specifically its relation to false positive rate (FPR) and true positive rate (TPR).

The full dynamic knowledge graph, along with the data required for feature preparation and evaluation, was processed on an Intel Xeon Gold 6130 CPU with 1 TiB of RAM. However, it is not strictly necessary to have 1 TiB of RAM for this process with the relatively small concept list of 37,960; the high memory capacity was utilized for efficiency and to handle additional operations concurrently. The final domain knowledge graph in this work occupies approximately 23.12 GB of storage. It is worth noting that knowledge graphs built from larger concept lists will require more memory, as the data size and complexity increase.

The neural network training was conducted on a standard single GPU (Nvidia Quadro RTX 6000), with each training run for different impact ranges taking approximately 1.5 hours. For benchmarking, all models -- except the transformer model -- were run on a standard CPU (Intel Xeon Gold 6130) with memory usage below 15 GB. Each benchmarking task took roughly one hour to complete. The transformer model, however, was run on a single GPU (Nvidia Quadro RTX 6000), taking approximately 3 hours per task.

\section{Individual feature's predictive ability} \label{appD} 
\begin{figure}[!t]
    \centering
    \includegraphics[width=0.98\linewidth]{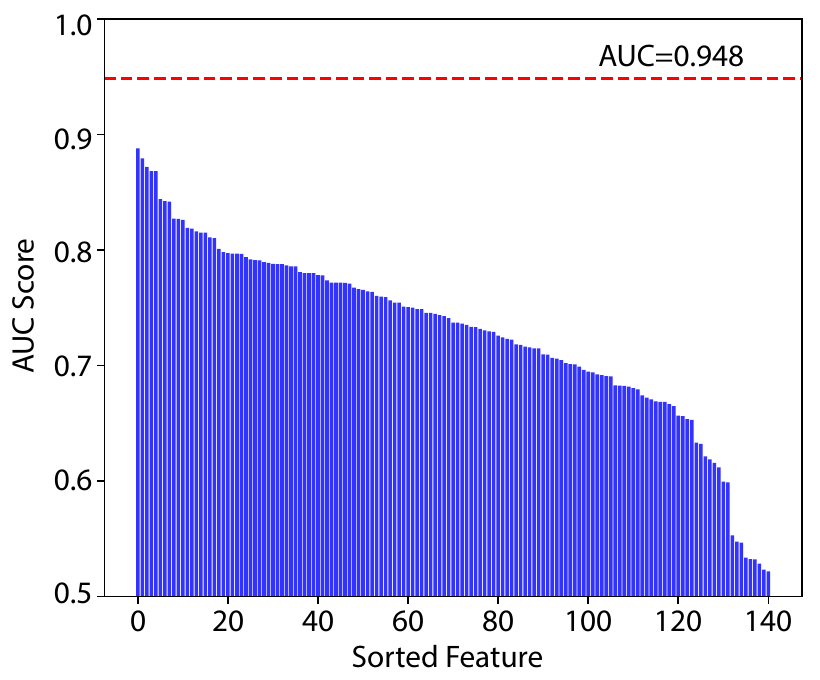}
    \caption{\textbf{Neural network performance across individual features.} The highest-performing four features are the Simpson similarity coefficient for the unconnected pair (\(u\), \(v\)) across the years \(y\), \(y-1\), and \(y-2\), and the cosine similarity coefficient for unconnected pairs (\(u\), \(v\)) in year \(y\) (i.e., \(y\)=2016), with AUC scores of 0.8880, 0.8795, 0.8720, and 0.8683, respectively. In contrast, the lowest predictive three features are the average total citation count up to year \(y\) for vertex \(v\), and the total citation count for the pair (\(u\), \(v\)) up to years \(y-1\) and \(y-2\), with AUC scores of 0.5219, 0.5234, and 0.5285. Using all 141 features together leads to a significant improvement in the AUC score to 0.948, showing that the combination of all features works better.} 
    \label{fig:NNsingle}
\end{figure}

In Fig. \ref{fig:auc_results} (a), we observe an AUC score of 0.948 for the 2019 evaluation dataset with the neural network that uses all 141 features, trained on 2016 dataset and impact range $IR=100$. To explore the predictive ability of individual features, we trained separate neural networks on each feature using the same 2016 dataset, and then applied the 2019 evaluation dataset to these models. This resulted in 141 individual predictions, each from a network trained on a single feature. The features were ranked by their impact predictions, shown in the Fig.~\ref{fig:NNsingle}.
Details of the features are shown in Tables~\ref{tab:features} and ~\ref{tab:features}, and the corresponding documentation and source code are available at \href{https://github.com/artificial-scientist-lab/Impact4Cast}{GitHub: Impact4Cast}.

\newpage
\renewcommand{\arraystretch}{1.5}
\begin{table*}[htbp]
\centering
\caption{Detailed description of all customized features for an unconnected concept pair $(u, v)$ for the year $y$}
\label{tab:features}
\begin{tabular}{c|c|p{13cm}}
\hline 
Type & Feature Index & Description  \\\hline \hline 
\multirow{6}{4em}{node feature} & 0-5 &Number of neighbors for each node ($u$ or $v$) until the year $y$, $y{-}1$, $y{-}2$\par denoted as $N_{u,y}$, $N_{v,y}$, $N_{u,y-1}$, $N_{v,y-1}$, $N_{u,y-2}$, and $N_{v,y-2}$, ordered as indices 0–5. \\  \cline{2-3}

& 6-7 & Number of new neighbors for each node ($u$ or $v$) between year $y{-}1$ and $y$\par i.e., $N_{u,y}{-}N_{u,y-1}$ and $N_{v,y}{-}N_{v,y-1}$.\\  \cline{2-3}

& 8-9 & Number of new neighbors for each node ($u$ or $v$) between year $y{-}2$ and $y$\par i.e., $N_{u,y}{-}N_{u,y-2}$ and $N_{v,y}{-}N_{v,y-2}$.\\  \cline{2-3}

& 10-11 & Rank of the number of new neighbors for each node ($u$ or $v$) between year $y{-}1$ and $y$ \par i.e., rank($N_{u,y}{-}N_{u,y-1}$) and rank($N_{v,y}{-}N_{v,y-1}$). \\  \cline{2-3}

& 12-13 & Rank of the number of new neighbors for each node ($u$ or $v$) between year $y{-}2$ and $y$\par i.e., rank($N_{u,y}{-}N_{u,y-1}$) and rank($N_{v,y}{-}N_{v,y-2}$). \\\cline{2-3}

& 14-19 & PageRank scores of each node ($u$ or $v$) until the year $y$, $y{-}1$, $y{-}2$\par denoted and ordered as $\mathrm{PR}_{u,y}$, $\mathrm{PR}_{v,y}$, $\mathrm{PR}_{u,y-1}$, $\mathrm{PR}_{v,y-1}$, $\mathrm{PR}_{u,y-2}$ and $\mathrm{PR}_{v,y-2}$. \\\hline 

\multirow{15}{4em}{node citation feature} & 20-25 & Yearly citation for each node ($u$ or $v$) in year $y$, $y{-}1$, $y{-}2$\par denoted and ordered as $\mathrm{C}_{u,y}$, $\mathrm{C}_{v,y}$, $\mathrm{C}_{u,y-1}$, $\mathrm{C}_{v,y-1}$, $\mathrm{C}_{u,y-2}$ and $\mathrm{C}_{v,y-2}$.  \\ \cline{2-3}

& 26-31 &Total citation for each node ($u$ or $v$) since the first publication to the year $y$, $y{-}1$, and $y{-}2$\par denoted and ordered as $\mathrm{Ct}_{u,y}$, $\mathrm{Ct}_{v,y}$, $\mathrm{Ct}_{u,y-1}$, $\mathrm{Ct}_{v,y-1}$, $\mathrm{Ct}_{u,y-2}$ and $\mathrm{Ct}_{v,y-2}$. \\ \cline{2-3}

& 32-37 & Total citations for each node ($u$ or $v$) in the last 3 years ending in the year $y$, $y{-}1$, and $y{-}2$\par
denoted and ordered as $\mathrm{Ct}^{\Delta 3}_{u,y}$, $\mathrm{Ct}^{\Delta 3}_{v,y}$, $\mathrm{Ct}^{\Delta 3}_{u,y{-}1}$, $\mathrm{Ct}^{\Delta 3}_{v,y{-}1}$, $\mathrm{Ct}^{\Delta 3}_{u,y{-}2}$, and $\mathrm{Ct}^{\Delta 3}_{v,y{-}2}$. \\ \cline{2-3}

& 38-43 &Number of papers mentioning node $u$ from the first publication to the year $y$, $y{-}1$, and $y{-}2$, similar for node $v$; denoted and ordered as $\mathrm{Pn}_{u,y}$, $\mathrm{Pn}_{v,y}$, $\mathrm{Pn}_{u,y-1}$, $\mathrm{Pn}_{v,y-1}$, $\mathrm{Pn}_{u,y-2}$, and $\mathrm{Pn}_{v,y-2}$\\ \cline{2-3}

& 44-49 & Average yearly citations for each node ($u$ or $v$) in the year $y$, $y{-}1$, $y{-}2$\par denoted and ordered as $\mathrm{Cm}_{u,y}$, $\mathrm{Cm}_{v,y}$, $\mathrm{Cm}_{u,y-1}$, $\mathrm{Cm}_{v,y-1}$, $\mathrm{Cm}_{u,y-2}$ and $\mathrm{Cm}_{v,y-2}$\par e.g., $\mathrm{Cm}_{u,y}=\mathrm{C}_{u,y}/\mathrm{Pn}_{u,y}$\\ \cline{2-3}

& 50-55 & Average total citations for each node ($u$ or $v$) since the first publications to the years $y$, $y{-}1$, $y{-}2$; denoted and ordered as $\mathrm{Ctm}_{u,y}$, $\mathrm{Ctm}_{v,y}$, $\mathrm{Ctm}_{u,y-1}$, $\mathrm{Ctm}_{v,y-1}$, $\mathrm{Ctm}_{u,y-2}$ and $\mathrm{Ctm}_{v,y-2}$; e.g., $\mathrm{Ctm}_{u,y}=\mathrm{Ct}_{u,y}/\mathrm{Pn}_{u,y}$\\\cline{2-3}

& 56-61 & Average total citations for each node ($u$ or $v$) in the last 3 years ending in the year $y$, $y{-}1$, and $y{-}2$; denoted and ordered as $\mathrm{Ctm}^{\Delta 3}_{u,y}$, $\mathrm{Ctm}^{\Delta 3}_{v,y}$, $\mathrm{Ctm}^{\Delta 3}_{u,y-1}$, $\mathrm{Ctm}^{\Delta 3}_{v,y-1}$, $\mathrm{Ctm}^{\Delta 3}_{u,y-2}$ and $\mathrm{Ctm}^{\Delta 3}_{v,y-2}$; e.g., $\mathrm{Ctm}^{\Delta 3}_{u,y}=\mathrm{Ct}^{\Delta 3}_{u,y}/\mathrm{Pn}_{u,y}$\\\cline{2-3}

& 62-63 & New citations for each node ($u$ or $v$) between years $y{-}1$ and $y$\par
i.e., $\mathrm{Ct}_{u,y}{-}\mathrm{Ct}_{u,y-1}$ and $\mathrm{Ct}_{v,y}{-}\mathrm{Ct}_{v,y-1}$. \\ \cline{2-3}

& 64-65 & New citations for each node ($u$ or $v$) between years $y{-}2$ and $y$\par
i.e., $\mathrm{Ct}_{u,y}{-}\mathrm{Ct}_{u,y-2}$ and $\mathrm{Ct}_{v,y}{-}\mathrm{Ct}_{v,y-2}$. \\ \cline{2-3}

& 66-67 & Rank of the new citations for each node ($u$ or $v$) between years $y{-}1$ and $y$\par i.e., rank($\mathrm{C}_{u,y}{-}\mathrm{C}_{u,y-1}$) and rank($\mathrm{C}_{v,y}{-}\mathrm{C}_{v,y-1}$). \\ \cline{2-3}

& 68-69 & Rank of the new citations for each node ($u$ or $v$) between years $y{-}2$ and $y$\par i.e., rank($\mathrm{C}_{u,y}{-}\mathrm{C}_{u,y-2}$) and rank($\mathrm{C}_{v,y}{-}\mathrm{C}_{v,y-2}$). \\ \cline{2-3}

& 70-71 & Number of papers mentioning nodes $u$ between years $y{-}1$ and $y$, similar for node $v$ \par i.e., $\mathrm{PR}_{u,y}-\mathrm{PR}_{u,y-1}$ and $\mathrm{PR}_{v,y}-\mathrm{PR}_{v,y-1}$ \\ \cline{2-3}

& 72-73  & Number of papers mentioning nodes $u$ between years $y{-}2$ and $y$, similar for node $v$ \par i.e., $\mathrm{PR}_{u,y}-\mathrm{PR}_{u,y-2}$ and $\mathrm{PR}_{v,y}-\mathrm{PR}_{v,y-2}$ \\ \cline{2-3}

& 74-75 & Rank of the number of papers mentioning nodes $u$ between years $y{-}1$ and $y$, similar for node $v$; i.e., rank($\mathrm{PR}_{u,y}-\mathrm{PR}_{u,y-1}$) and rank($\mathrm{PR}_{v,y}-\mathrm{PR}_{v,y-1}$) \\\cline{2-3}
& 76-77 & Number of papers mentioning nodes $u$ between years $y{-}2$ and $y$, similar for node $v$ \par i.e., rank($\mathrm{PR}_{u,y}-\mathrm{PR}_{u,y-2}$) and rank($\mathrm{PR}_{v,y}-\mathrm{PR}_{v,y-2}$) 
\\\hline \hline 
\end{tabular}
\end{table*}

\begin{table*}[htbp]
\centering
\caption{continued from previous page for Table.~\ref{tab:features}}
\label{tab:features2}
\begin{tabular}{c|c|p{13cm}}
\hline
Type & Feature Index & Description \\
\hline \hline
\multirow{7}{4em}{pair feature} & 78-80 & Number of shared neighbors between nodes $u$ and $v$ until the year $y$, $y{-}1$, $y{-}2$, denoted and ordered as $\mathrm{Ns}_{y}$, $\mathrm{Ns}_{y-1}$ and $\mathrm{Ns}_{y-2}$; e.g., $\mathrm{Ns}_{y}=\mathrm{N}_{u,y} \cap \mathrm{N}_{v,y}$  \\ \cline{2-3} 

& 81-83 & Geometric similarity coefficient for the pair $(u, v)$ for the year $y$, $y{-}1$, and $y{-}2$\par denoted and ordered as $\mathrm{Geo}_{y}$, $\mathrm{Geo}_{y-1}$, and $\mathrm{Geo}_{y-2}$; e.g., $\mathrm{Geo}_{y} = \mathrm{Ns}_{y}^{2}/(\mathrm{N}_{u,y}\times \mathrm{N}_{v,y})$. \\ \cline{2-3}

& 84-86 & Cosine similarity coefficient for the pair $(u, v)$ for the year $y$, $y{-}1$, $y{-}2$ \par
denoted and ordered as $\mathrm{Cos}_{y}$, $\mathrm{Cos}_{y-1}$, and $\mathrm{Cos}_{y-2}$; e.g.,  $\mathrm{Cos}_{y} = \sqrt{\mathrm{Geo}_{y}}$.\\\cline{2-3}

& 87-89 & Simpson coefficient for the pair $(u, v)$ for the year $y$, $y{-}1$, $y{-}2$ \par
denoted and ordered as $\mathrm{Sim}_{y}$, $\mathrm{Sim}_{y-1}$, and $\mathrm{Sim}_{y-2}$; e.g., $\mathrm{Sim}_{y} = \mathrm{Ns}_{y}/\min(\mathrm{N}_{u,y}, \mathrm{N}_{v,y})$.\\\cline{2-3}

& 90-92 & Preferential attachment coefficient for the pair $(u, v)$ for the year $y$, $y{-}1$, $y{-}2$\par
denoted and ordered as $\mathrm{Pre}_{y}$, $\mathrm{Pre}_{y-1}$, and $\mathrm{Pre}_{y-2}$; e.g.,  $\mathrm{Pre}_{y} =\mathrm{N}_{u,y}\times \mathrm{N}_{v,y}$.\\\cline{2-3}

& 93-95 &Sørensen–Dice coefficient for the pair $(u, v)$ for the year $y$, $y{-}1$, $y{-}2$\par
denoted and ordered as $\mathrm{Sor}_{y}$, $\mathrm{Sor}_{y-1}$, and $\mathrm{Sor}_{y-2}$; e.g., $\mathrm{Sor}_{y} = 2\mathrm{Ns}_{y}/(\mathrm{N}_{u,y}+\mathrm{N}_{v,y}$).\\\cline{2-3}

& 96-98 & Jaccard coefficient for the pair $(u, v)$ for the year $y$, $y{-}1$, $y{-}2$\par
denoted and ordered as $\mathrm{Jac}_{y}$, $\mathrm{Jac}_{y-1}$, and $\mathrm{Jac}_{y-2}$; e.g., $\mathrm{Jac}_{y} = \mathrm{Ns}_{y}/(\mathrm{N}_{u,y}+\mathrm{N}_{v,y}-\mathrm{Ns}_{y})$.\\\hline

\multirow{14}{4em}{pair citation feature} & 99-101 & Ratio of the sum of citations received by nodes $u$ and $v$ until the year $y$ to the total number of papers mentioning either concept; denoted and ordered as $\mathrm{r1}_{y}$, $\mathrm{r1}_{y-1}$, and $\mathrm{r1}_{y-2}$; e.g., $\mathrm{r1}_{y}=(\mathrm{Ct}_{u,y}$ + $\mathrm{Ct}_{v,y}) / (\mathrm{Pn}_{u,y}+\mathrm{Pn}_{v,y})$.\\\cline{2-3}

& 102-104 & Ratio of the product of citations received by nodes $u$ and $v$ until the year $y$ to the total number of papers mentioning either concept; denoted and ordered as $\mathrm{r2}_{y}$, $\mathrm{r2}_{y-1}$, and $\mathrm{r2}_{y-2}$; e.g., $\mathrm{r2}_{y}=(\mathrm{Ct}_{u,y} \times \mathrm{Ct}_{v,y}) / (\mathrm{Pn}_{u,y}+\mathrm{Pn}_{v,y})$.\\\cline{2-3}

& 105-107 & Sum of average citations received by nodes $u$ and $v$ in the year $y$, $y{-}1$, and $y{-}2$\par denoted and ordered as $\mathrm{s}_{y}$, $\mathrm{s}_{y-1}$, and $\mathrm{s}_{y-2}$; e.g., $\mathrm{s}_{y}=\mathrm{Cm}_{u,y}+\mathrm{Cm}_{v,y}$.\\\cline{2-3}

& 108-110 & Sum of average total citations received by nodes $u$ and $v$ from the first publication to the year $y$, $y{-}1$, and $y{-}2$; denoted and ordered as $\mathrm{st}_{y}$, $\mathrm{st}_{y-1}$, and $\mathrm{st}_{y-2}$; e.g., $\mathrm{st}_{y}=\mathrm{Ctm}_{u,y}+\mathrm{Ctm}_{v,y}$.\\\cline{2-3}

& 111-113 & Sum of the total citations received by nodes $u$ and $v$ in the last 3 years ending in the year $y$, $y{-}1$, and $y{-}2$; denoted and ordered as $\mathrm{st}^{\Delta 3}_{y}$, $\mathrm{st}^{\Delta 3}_{y-1}$, and $\mathrm{st}^{\Delta 3}_{y-2}$; e.g., $\mathrm{st}^{\Delta 3}_{y}=\mathrm{Ct}^{\Delta 3}_{u,y}+\mathrm{Ct}^{\Delta 3}_{v,y}$. \\ \cline{2-3}

& 114-116 & Sum of average total citations received by nodes $u$ and $v$ in the last 3 years ending in the year $y$, $y{-}1$, and $y{-}2$; denoted and ordered as $\mathrm{stm}^{\Delta 3}_{y}$, $\mathrm{stm}^{\Delta 3}_{y-1}$, and $\mathrm{stm}^{\Delta 3}_{y-2}$\par e.g., $\mathrm{stm}^{\Delta 3}_{y}=\mathrm{Ctm}^{\Delta 3}_{u,y}+\mathrm{Ctm}^{\Delta 3}_{v,y}$. \\\cline{2-3}

& 117-119 & Minimum number of citations received by either node $u$ or $v$ in years $y$, $y{-}1$, and $y{-}2$; denoted and ordered as $\mathrm{minC}_{y}$, $\mathrm{minC}_{y-1}$, and $\mathrm{minC}_{y-2}$; e.g., $\mathrm{minC}_{y}=\min(\mathrm{C}_{u,y}, \mathrm{C}_{v,y})$. \\ \cline{2-3}

& 120-122 & Maximum number of citations received by either node $u$ or $v$ in years $y$, $y{-}1$, and $y{-}2$; denoted and ordered as $\mathrm{maxC}_{y}$, $\mathrm{maxC}_{y-1}$, and $\mathrm{maxC}_{y-2}$; e.g., $\mathrm{maxC}_{y}=\max(\mathrm{C}_{u,y}, \mathrm{C}_{v,y})$. \\ \cline{2-3}

& 123-125 & Minimum number of total citations received by nodes $u$ and $v$ from the first publication to the year $y$, $y{-}1$, and $y{-}2$; denoted and ordered as $\mathrm{minCt}_{y}$, $\mathrm{minCt}_{y-1}$, and $\mathrm{minCt}_{y-2}$;\par e.g., $\mathrm{minCt}_{y}= \min(\mathrm{Ct}_{u,y}, \mathrm{Ct}_{v,y})$.\\\cline{2-3}

& 126-128 & Maximum number of total citations received by nodes $u$ and $v$ from the first publication to the year $y$, $y{-}1$, and $y{-}2$; denoted and ordered as $\mathrm{maxCt}_{y}$, $\mathrm{maxCt}_{y-1}$, and $\mathrm{maxCt}_{y-2}$;\par e.g., $\mathrm{maxCt}_{y}=\max(\mathrm{Ct}_{u,y}, \mathrm{Ct}_{v,y})$.\\\cline{2-3}

& 129-131 & Minimum number of total citations received by nodes $u$ and $v$ in the last 3 years ending in the year $y$, $y{-}1$, and $y{-}2$; denoted and ordered as $\mathrm{minCt}^{\Delta 3}_{y}$, $\mathrm{minCt}^{\Delta 3}_{y-1}$, and $\mathrm{minCt}^{\Delta 3}_{y-2}$;\par e.g., $\mathrm{minCt}^{\Delta 3}_{y}= \min(\mathrm{Ct}^{\Delta 3}_{u,y}, \mathrm{Ct}^{\Delta 3}_{v,y})$.\\\cline{2-3}

& 132-134 & Maximum number of total citations received by nodes $u$ and $v$ in the last 3 years ending in the year $y$, $y{-}1$, and $y{-}2$; denoted and ordered as $\mathrm{maxCt}^{\Delta 3}_{y}$, $\mathrm{maxCt}^{\Delta 3}_{y-1}$, and $\mathrm{maxCt}^{\Delta 3}_{y-2}$;\par e.g., $\mathrm{maxCt}^{\Delta 3}_{y}= \max(\mathrm{Ct}^{\Delta 3}_{u,y}, \mathrm{Ct}^{\Delta 3}_{v,y})$.\\\cline{2-3}

& 135-137 & Minimum number of papers mentioning the node $u$ or node $v$ from the first publication to the year $y$, $y{-}1$, and $y{-}2$; denoted and ordered as $\mathrm{minPn}_{y}$, $\mathrm{minPn}_{y-1}$ and $\mathrm{minPn}_{y-2}$;\par e.g., $\mathrm{minPn}_{y}= \min(\mathrm{Pn}_{u,y}, \mathrm{Pn}_{v,y})$.\\ \cline{2-3}

& 138-140 & Maximum number of papers mentioning the node $u$ or node $v$ from the first publication to the year $y$, $y{-}1$, and $y{-}2$; denoted and ordered as $\mathrm{maxPn}_{y}$, $\mathrm{maxPn}_{y-1}$ and $\mathrm{maxPn}_{y-2}$;\par e.g., $\mathrm{maxPn}_{y}= \max(\mathrm{Pn}_{u,y}, \mathrm{Pn}_{v,y})$.\\ \hline 

\end{tabular}
\end{table*}

\clearpage
\bibliography{ref}

\end{document}